\documentclass[3p,times,twocolumn]{elsarticle}

\usepackage{ecrc}

\volume{00}

\firstpage{1}

\journalname{Nuclear Physics B Proceedings Supplement}

\runauth{G.W.~Sullivan}

\jid{nuphbp}

\jnltitlelogo{Nuclear Physics B Proceedings Supplement}

\usepackage{amssymb}

\usepackage[figuresright]{rotating}

\begin{document}

\begin{frontmatter}



\dochead{}

\title{Results from the IceCube Experiment}

\author[label1]{G.W.~Sullivan for the IceCube Collaboration}
\address[label1]{Department of Physics, University of Maryland, College Park, MD 20742, USA}

\begin{abstract}
The IceCube detector, which completed construction in December 2010, is the first km$^3$ scale instrument to become operational with the primary mission of observing high energy neutrinos from astrophysical sources to elucidate our understanding of the source of cosmic rays and the astrophysical mechanisms that produce them. Here we report on results  from the partially completed detector on searches for diffuse astrophysical neutrinos,  astrophysical point sources of neutrinos and neutrino emission coincident with gamma ray bursts. We also present results on the observation of neutrino oscillations and discuss the potential program for neutrino oscillation studies. 
\end{abstract}

\end{frontmatter}


\section{Introduction}
\label{sec:introduction}
It has now been over 100 years since the discovery  of charged particle cosmic rays by Victor Hess. Since then, much has been learned about them, including information on their spectrum and particle composition. The cosmic ray flux has been measured over more than 10 orders of magnitude in energy and has been observed to energies of  $10^{20}$ eV.  Yet, because the charged cosmic rays deflect in the magnetic fields and do not point to their origin, we still do not know where these charged particles are produced and what astrophysical processes are at work in accelerating them to such extreme energies. In recent years, observations of high energy gamma rays have provided a great deal of new information on the high energy behavior of many astrophysical sources that are candidate sources for cosmic rays. However, these measurements are limited in energy reach for extragalactic objects due to the absorption of gamma rays over several TeV in energy, and because there are electromagnetic processes that can contribute to their production, which would not accelerate the observed hadrons in cosmic rays.

Neutrinos provide a unique messenger particle for high energy astrophysical measurement since they neither deflect in magnetic fields nor interact with extragalactic background light at higher energies. This ``neutrino astronomy" will provide a unique window on the high energy universe and the physical mechanisms at work in high energy astrophysical objects. The observation of neutrinos from astrophysical regions would also provide unambiguous evidence for hadronic acceleration and the source of cosmic rays. 

\begin{figure}
\begin{center}
	\resizebox{\linewidth}{!}{\includegraphics{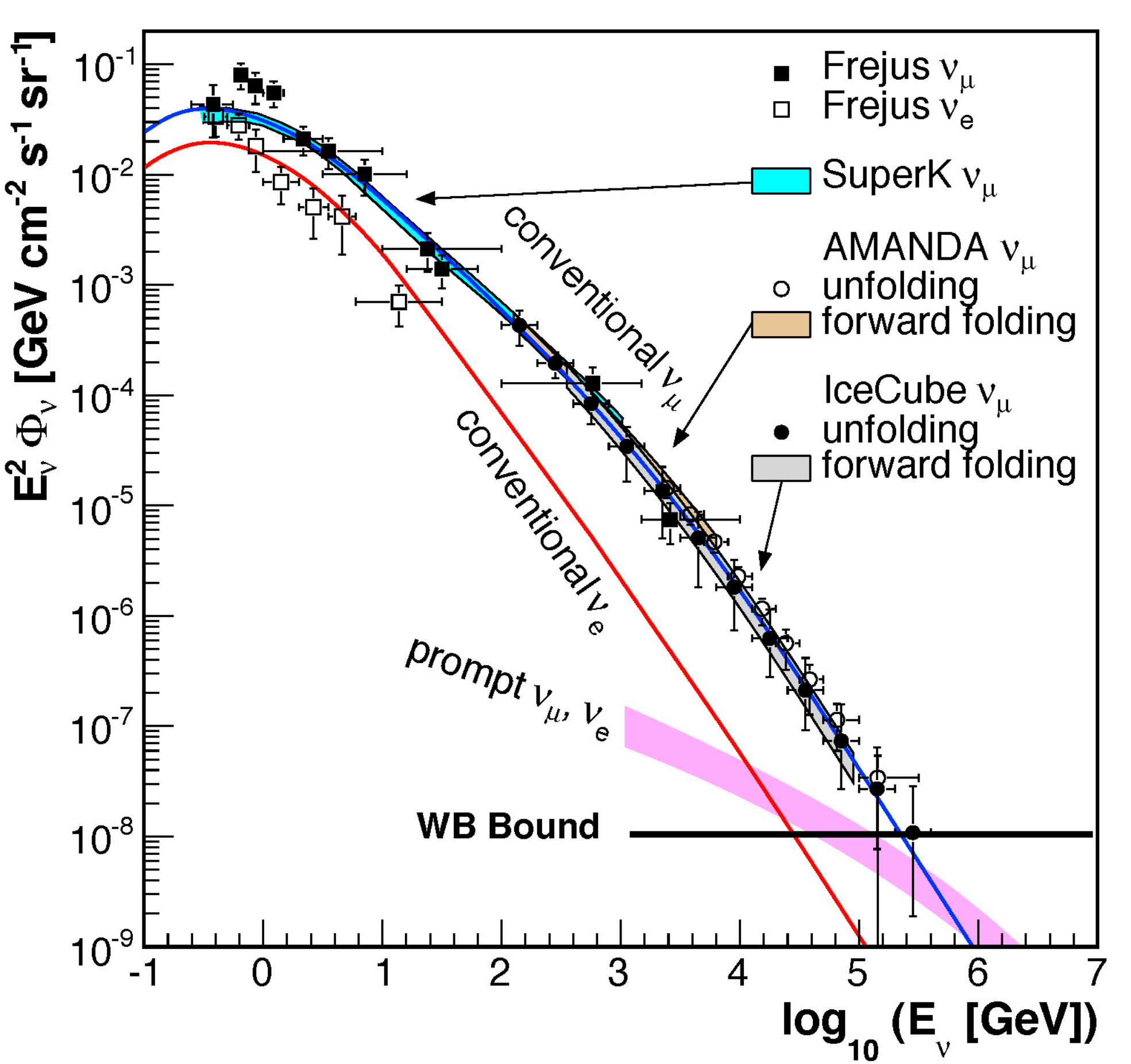}}
	\caption{The atmospheric neutrino spectrum as a function of energy. The symbols are various measurements and the curves are theoretical predictions. The prompt flux band represents the theoretical uncertainty. Also shown as a horizontal line is the Waxman-Bahcall (WB) upper bound, which defines the needed sensitivity and energy range of a neutrino telescope.}
	\label{fig:atmospheric}
\end{center}
\end{figure}

The very properties of neutrinos and their weak interaction that make them ideal messenger particles also make them difficult to detect with sufficient sensitivity. It has been a goal of the neutrino telescope community for decades~\cite{history} to develop an instrument of approximately 1 km$^3$ in size to reach the sensitivity of astrophysical interest \cite{halzen}. Figure~\ref{fig:atmospheric} shows the neutrino flux observed at earth comprised of atmospheric neutrinos that are produced in the cosmic ray interactions with the atmosphere. These atmospheric neutrinos provide a continuous source for measurements of neutrino properties while also representing an irreducible background in the search for astrophysical neutrinos. 

The IceCube detector is the first neutrino telescope on the scale of one cubic kilometer with the sensitivity to neutrino fluxes of astrophysical interest. Figure~\ref{fig:atmospheric} shows the falling spectrum of atmospheric neutrinos along with the Waxman-Bahcall (WB) upper bound \cite{WB} that sets the scale for the predicted flux and energy range of astrophysical neutrinos compatible with the known flux of high energy cosmic rays. The sensitivity of the partial IceCube detector (IC40 and IC59) has already reached this upper bound.

\section{Detector}
\label{sec:detector}

\begin{figure}
\begin{center}
	\resizebox{\linewidth}{!}{\includegraphics{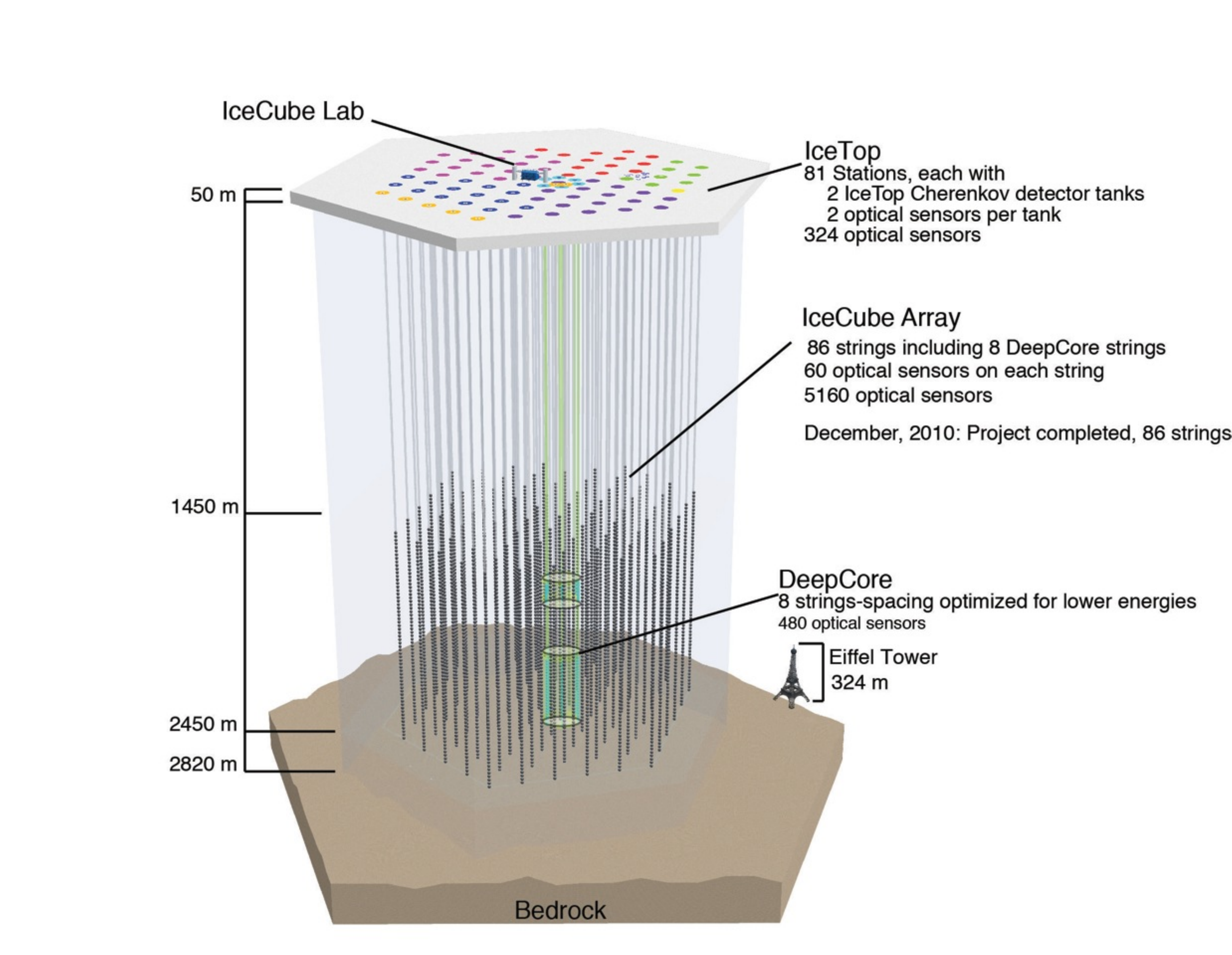}}
	\caption{The layout of the IceCube detector.}
	\label{fig:detector}
\end{center}
\end{figure}

The IceCube detector is a Cherenkov light tracking detector, which uses the deep clear ice at the South Pole as both the interaction and detector medium (see Figure~\ref{fig:detector}). The detector was deployed during the six austral summer seasons with completion in December 2010.   During the years of construction the detector was run in partial configurations denoted by the ``Dataset" labels shown in Table~\ref{tab:runs}. The ice volume is instrumented with 86 strings of 60 Digital Optical Modules (DOMs) between 1450m and 2450m depth. The DOM is a pressure vessel that contains a 25.4 cm diameter Hamamatsu photomultiplier, digitizing, timestamping, high voltage and calibration electronics. The DOMs communicate digitally over copper twisted-pair to the surface DAQ system in the IceCube Laboratory surface building located near the center of the array. Eight of the strings in the center of the array are deployed with a higher density of DOMs in the deep ice infilling the typical 125 meters string spacing. This inner core of high density DOMs comprises the DeepCore array, which was deployed to extend the low energy response of the detector. There are also 81 IceTop stations on the surface that form a high energy air-shower detector for the study of cosmic rays and added background rejection to the IceCube neutrino physics. The event readout trigger rate is approximately 2700 Hz and the raw data are processed by an online filter system to reduce the overall data rate to the approximately 100GB per day that can be transmitted north over the satellite. The online system also develops a near realtime neutrino stream for use as an active alert system. The overall rate of well reconstructed high energy ($>$ 100's GeV) atmospheric neutrinos is approximately 100 per day. 

IceCube reconstructs events based on the amplitude, spatial and time pattern of the light detected by the DOMs. A muon from a cosmic ray shower or charged current (CC) muon neutrino interaction produces a long track in the detector, while neutral current events and CC electron and tau neutrinos produce almost spherical ``cascade" signatures. The pointing resolution for muon neutrinos is on the order of one degree, better for higher energies, which includes the opening angle between the incoming neutrino and the muon from the CC interaction.
 
\begin{table}[h]
\begin{center}
	\begin{tabular}{|c|c|c|}
	\hline
	Dataset & Strings & Run Dates \\
	\hline \hline
	IC22 & 22 & May 2007 - Apr 2008 \\
	\hline
	IC40 & 40 & Apr 2008 - May 2009 \\
	\hline
	IC59 & 59 & May 2009 - May 2010 \\
	\hline
	IC79 & 79 & May 2010 - Apr 2011 \\
	\hline
	IC86-1 & 86 & Apr 2011 - May 2012 \\
	\hline
	IC86-2 & 86 & May 2012 - present \\
	\hline
	\end{tabular}
	\caption{IceCube run configurations and dates.}
	\label{tab:runs}
\end{center}
\end{table}

\section{Results}
\label{sec:results}

\subsection{Diffuse Astrophysical Neutrinos}

\begin{figure}
\begin{center}
	\resizebox{\linewidth}{!}{\includegraphics{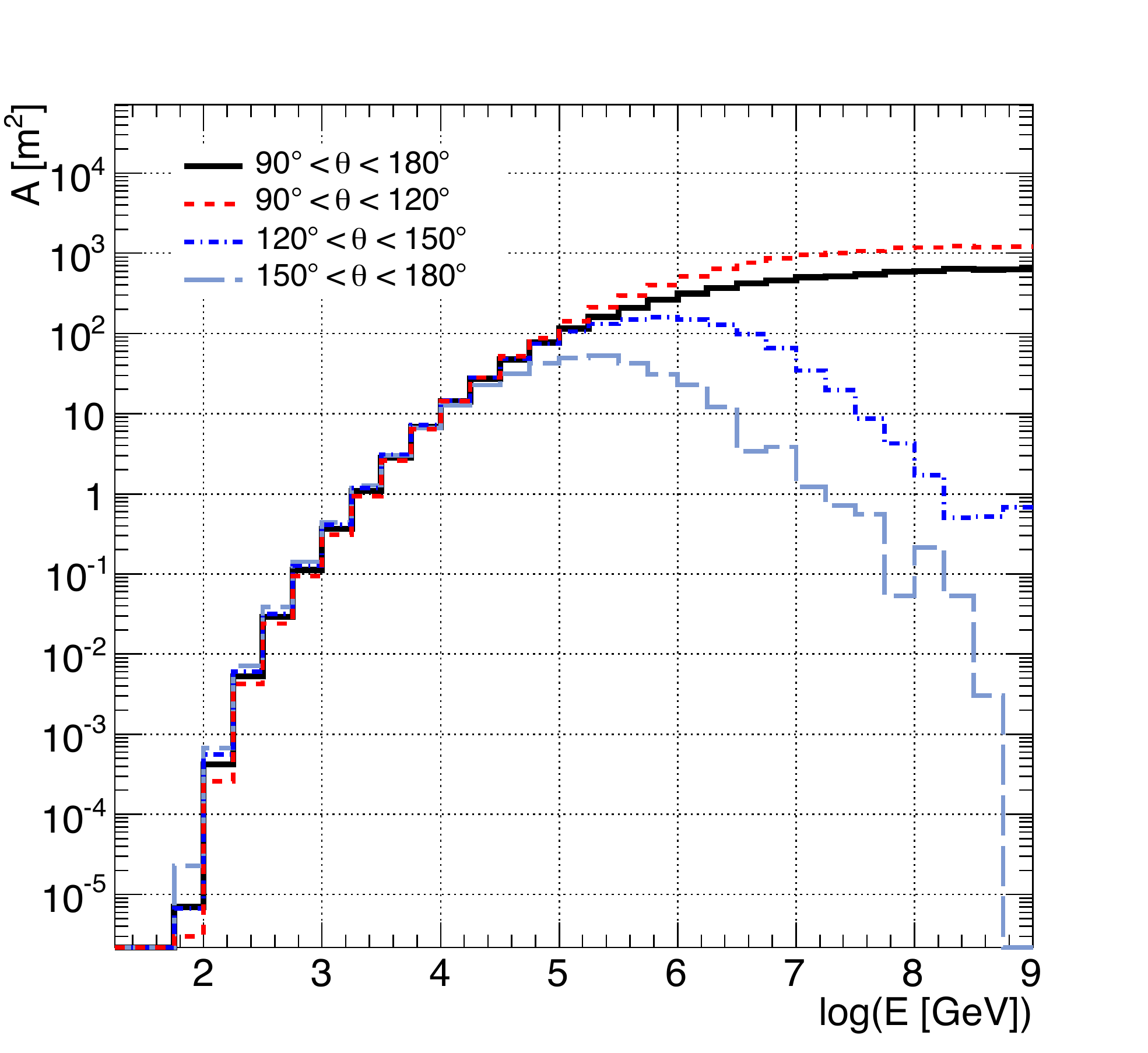}}
	\caption{The neutrino effective area versus neutrino energy for various zenith angle bands in the IC59 diffuse muon neutrino search.}
	\label{fig:area}
\end{center}
\end{figure}

We report here the preliminary result on the search for diffuse astrophysical muon neutrinos using the IC59 dataset. The diffuse astrophysical neutrino spectrum is expected to have an E$^{-2}$ spectrum, while the background atmospheric neutrinos have a softer spectrum of E$^{-3.7}$. This analysis searches for a diffuse flux of astrophysical muon neutrinos by looking for upward going tracks at energies above the rapidly falling atmospheric neutrino spectrum. After all cuts there are 21,943 events for 348 live-days. The neutrino purity of the sample (dominated by atmospheric neutrinos) is  99.8\% with efficiency of 12\% and 30\% for atmospheric and E$^{-2}$ spectra respectively. Figure~\ref{fig:area} shows the neutrino effective area as a function of neutrino energy for various zenith ranges.

\begin{figure}
\begin{center}
	\resizebox{\linewidth}{!}{\includegraphics{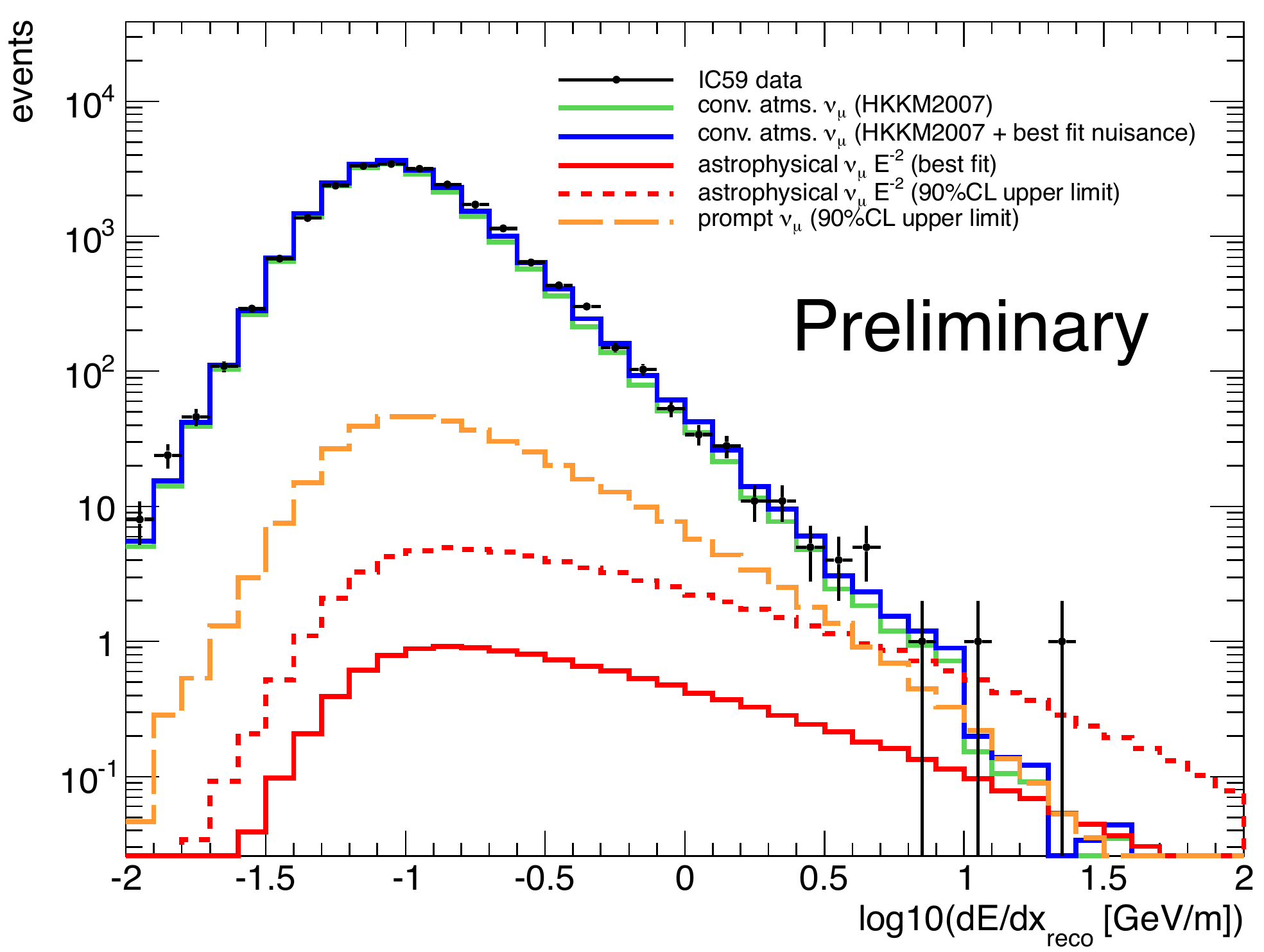}}
	\caption{The neutrino data and best fit spectra for atmospheric, prompt and astrophysical diffuse muon neutrinos in the IC59 analysis. Also shown are the 90\% CL upper limits for the prompt (gold long-dash) and E$^{-2}$ astrophysical (red short-dashed).} 
	\label{fig:diffuse}
\end{center}
\end{figure}

\begin{figure}
\begin{center}
	\resizebox{\linewidth}{!}{\includegraphics{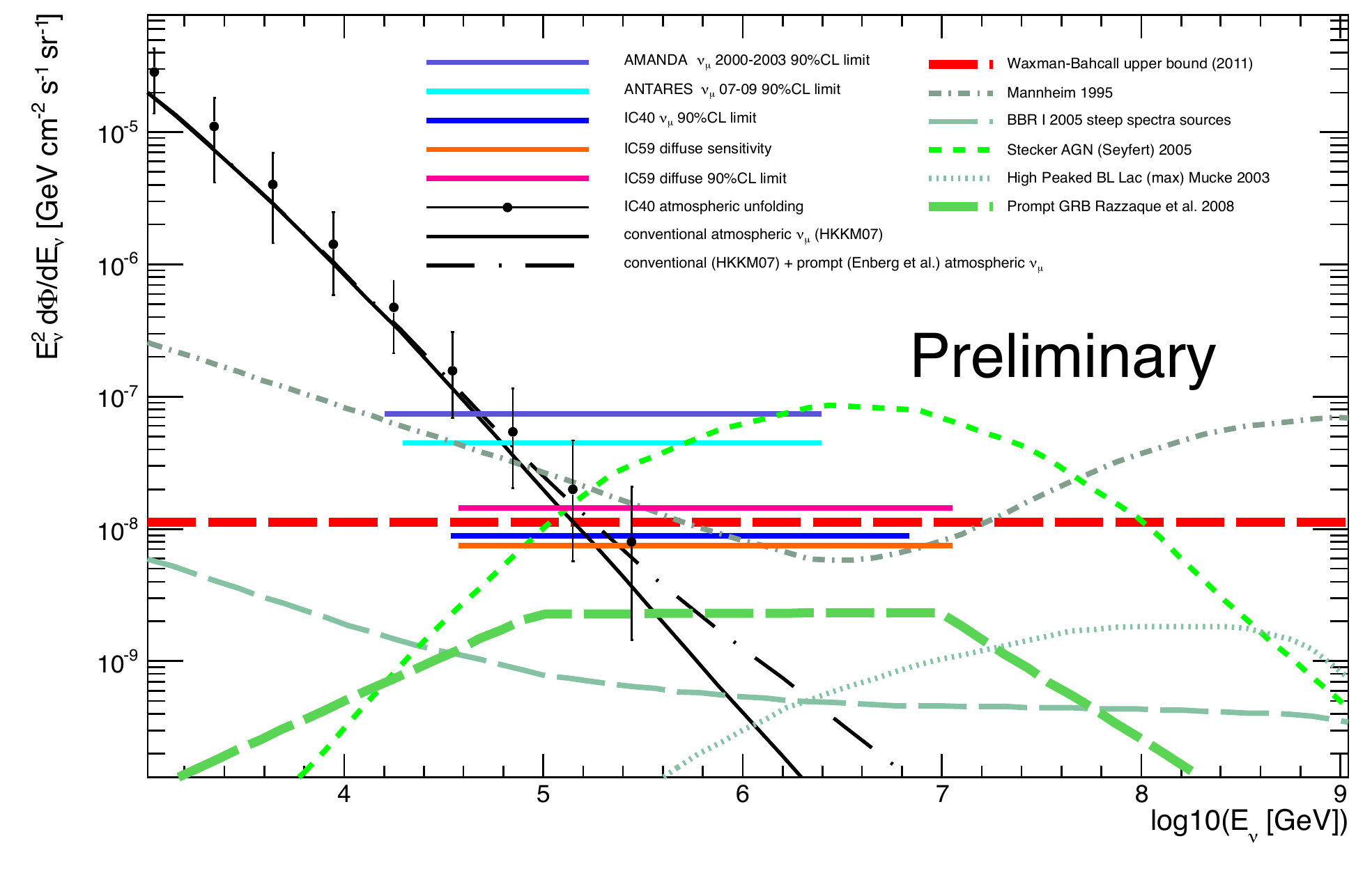}}
	\caption{Summary $\nu_\mu$ diffuse astrophysical limits.}
	\label{fig:diflimit}
\end{center}
\end{figure}

Shown in Figure~\ref{fig:diffuse} is the spectrum of events versus the energy deposition parameter for the muon neutrino data along with the best fit diffuse astrophysical flux (red solid). The best fit astrophysical flux is non-zero, but consistent with zero at less than 2 sigma. The 90\% CL flux upper limit (red dashed) is $\phi _{lim}$ E$^2 <  1.44 \times 10^{-8}$ GeV sr$^{-1}$ s$^{-1}$ cm$^{-2}$. The limit from this IC59 analysis is shown in Figure~\ref{fig:diflimit}.  In the figure the points represent the IC40 neutrino data. The bold horizontal dashed red line is the WB upper bound. Various models for astrophysical neutrino fluxes are shown in the broken lines while the limits from experiments are the the solid  horizontal lines from top to bottom, AMANDA (purple), ANTARES 07-08 (aqua), IC59 (pink), IC40 (blue), and IC59 sensitivity (orange). Note that while the sensitivity from this analysis is below the IC40 limit and the WB bound, the IC59 limit is above both because of the non-zero signal in the best fit.


\subsection{Astrophysical Point Sources}

\begin{figure}
\begin{center}
	\resizebox{\linewidth}{!}{\includegraphics{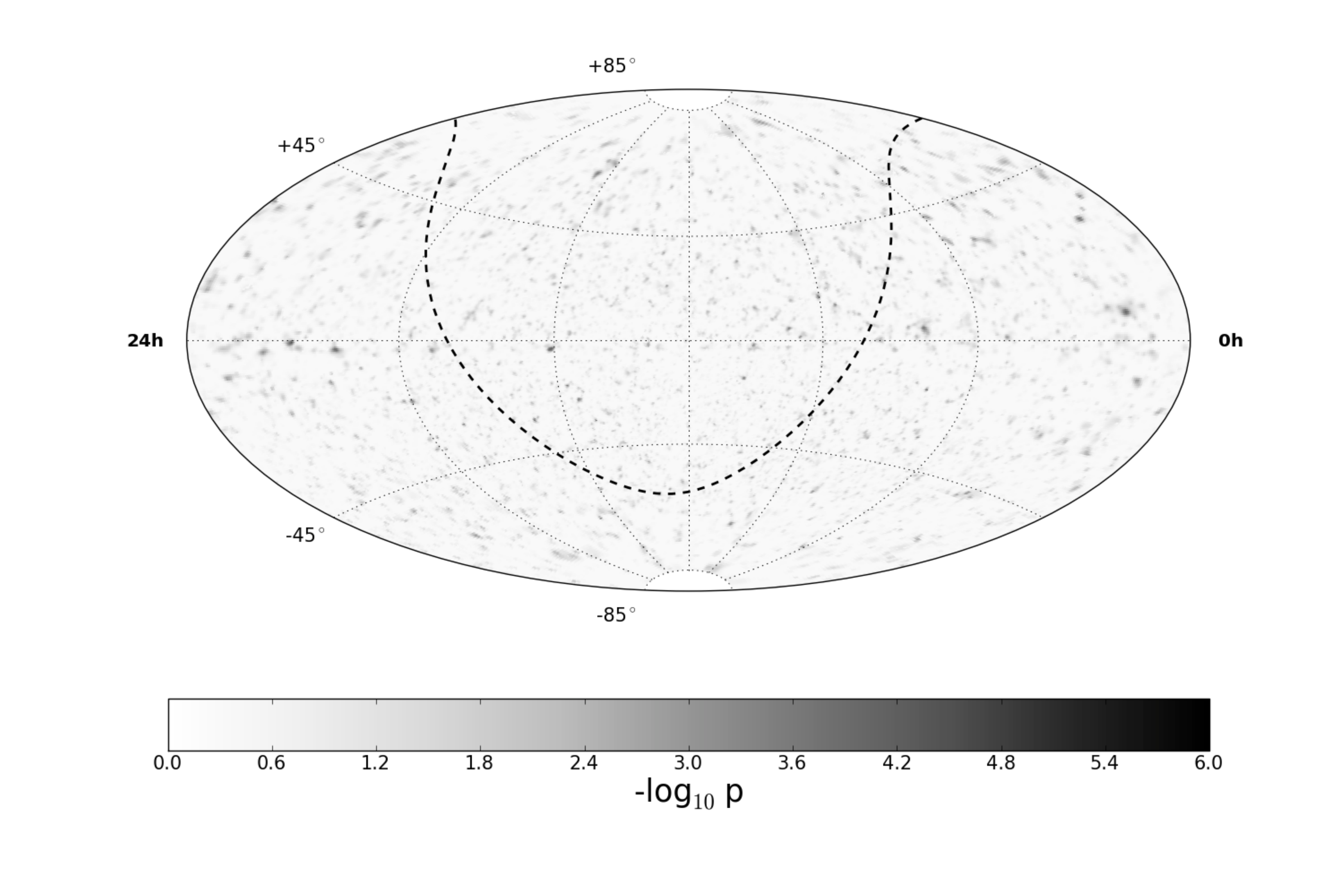}}
	\caption{IceCube IC40 + IC59 all sky point source significance map in equatorial coordinates. The p-value is not corrected for trials. Post trials, there is no significant region in the map.}
	\label{fig:skymap}
\end{center}
\end{figure}

\begin{figure}
\begin{center}
	\resizebox{\linewidth}{!}{\includegraphics{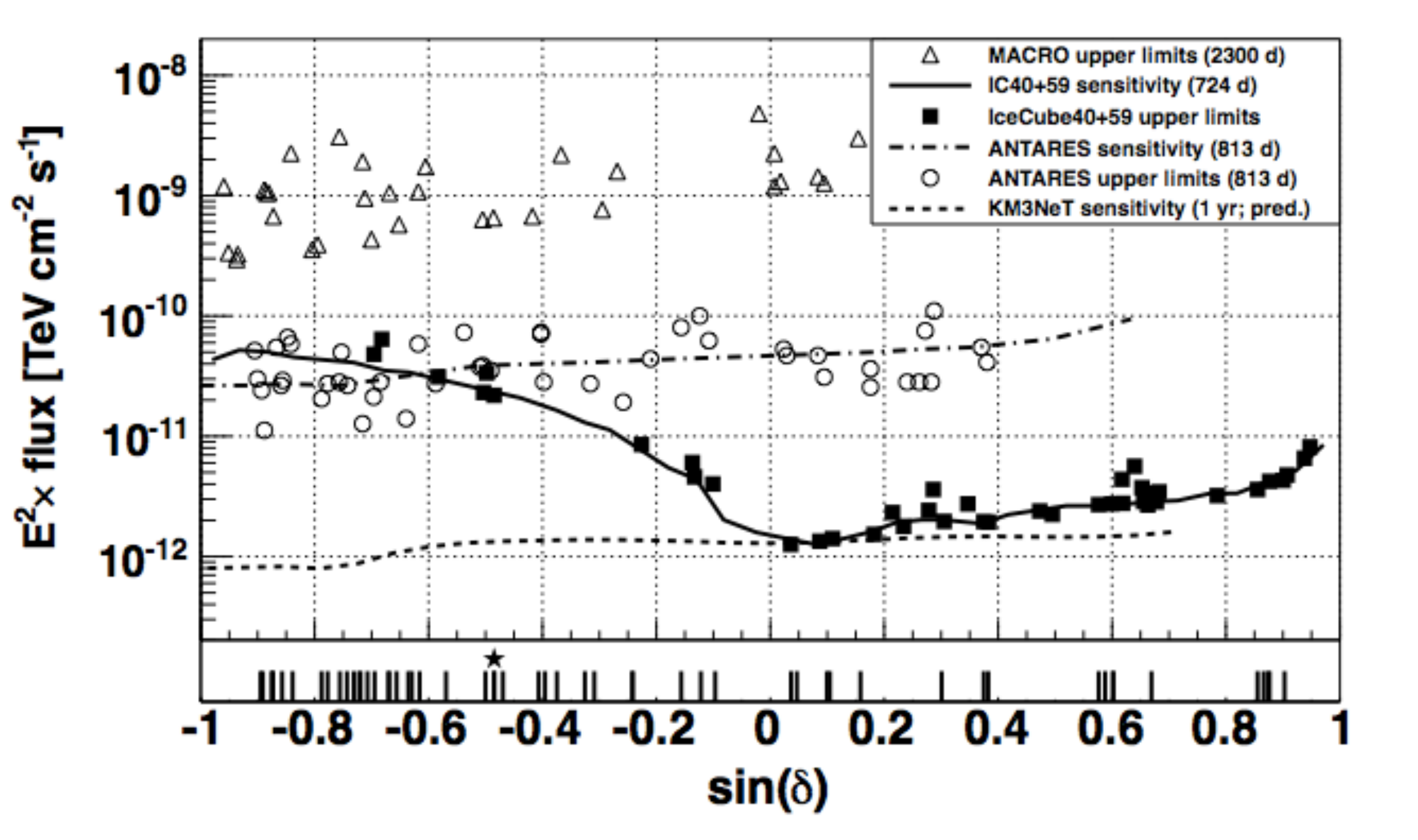}}
	\caption{90\% CL upper limits (symbols) and sensitivities (lines) for point sources with an E$^{-2}$ spectrum as function of declination. The vertical lines at the bottom represent the positions of known Galactic TeV gamma- ray emitters. The star marks the position of the Galactic center.}
	\label{fig:pslimit}
\end{center}
\end{figure}

An all sky time-independent point source search was performed using the IC40 and IC59 muon data sets. This analysis has a total livetime of 723 days, with on the order of 45,000 events from the northern hemisphere (upward going) and 65,000 events from the southern hemisphere (downward going). The events were used to create a likelihood map of the entire sky shown in Figure~\ref{fig:skymap}. The brightest spot in the sky has a post trials p-value of 67\% and is consistent with a background fluctuation. Figure~\ref{fig:pslimit} shows the 90\% CL upper limits (solid symbols) and sensitivities (solid line) for point sources with an E$^{-2}$ spectrum versus the source declination. Also shown are the MACRO and ANTARES limits, and the future KM3Net sensitivity. Notice that for IceCube, although the sensitivity to the southern hemisphere is reduced because of background from cosmic ray muons, the detector has sensitivity to the entire sky. Although no sources have yet been identified with the partial detector, we expect to reach the real region of interest in the next few years.

\subsection{Neutrino Emission from Gamma Ray Bursts}

Gamma Ray Bursts (GRBs) are brief (0.1s-100s)  intense flashes of gamma rays that have been detected by a number of gamma ray instruments over the decades. GRBs are now known to be of extragalactic origin with massive energy release, and have been proposed as possible candidate sources for the high energy cosmic rays~\cite{GRB}. In the GRB ÔfireballÕ model, cosmic-ray acceleration should be accompanied by neutrinos produced in the decay of charged pions created in interactions between the high-energy cosmic-ray protons and gamma rays. 

\begin{figure}
\begin{center}
	\resizebox{\linewidth}{!}{\includegraphics{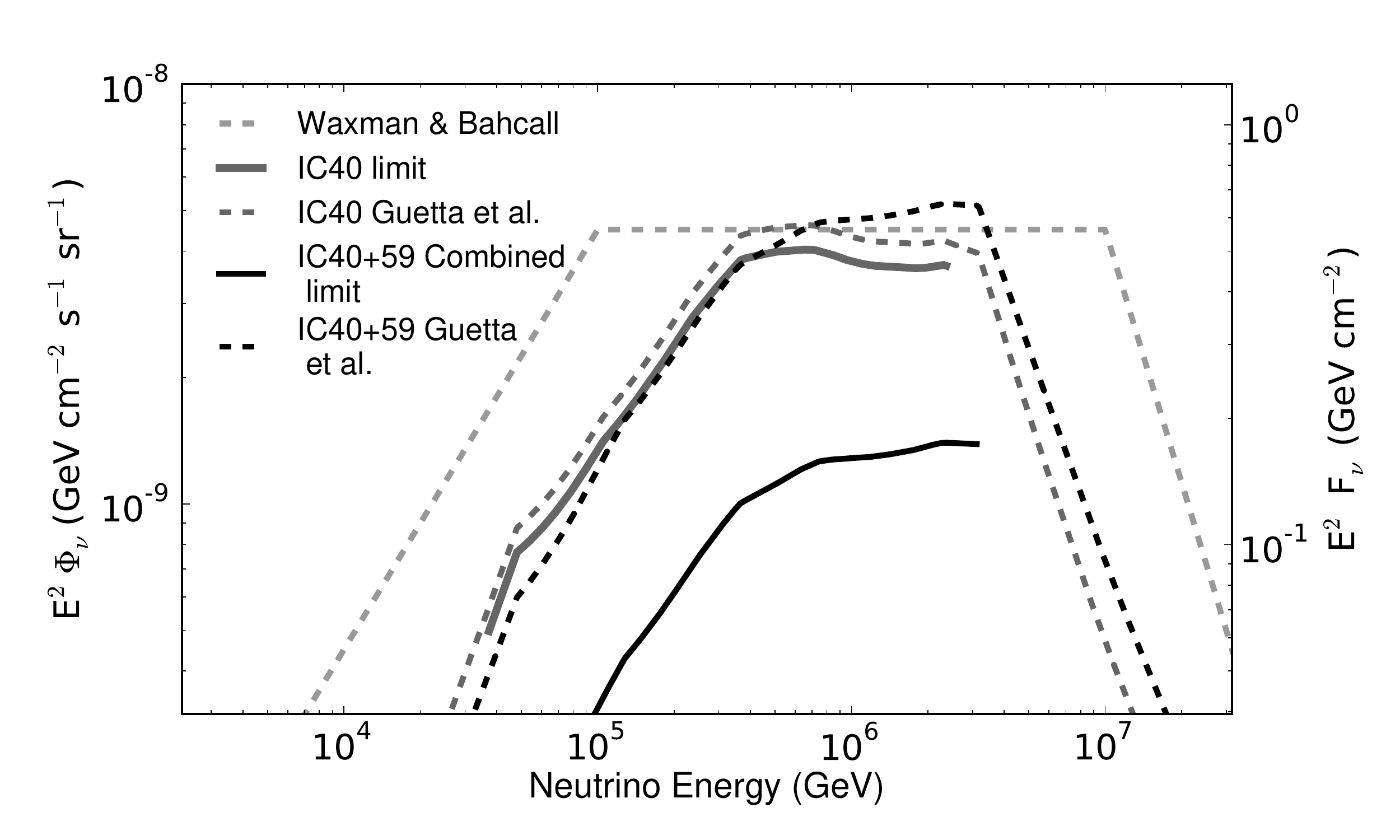}}
	\caption{90\% CL upper limits (solid) compared to summed flux predictions normalized to the gamma ray spectra (dashed) versus energy for the IC40 and IC40+IC59 data. Also shown in dashed is the generic WB model. The current limit is well below the prediction. Taken from~\cite{nature}. }
	\label{fig:grblimit}
\end{center}
\end{figure}

\begin{figure}
\begin{center}
	\resizebox{\linewidth}{!}{\includegraphics{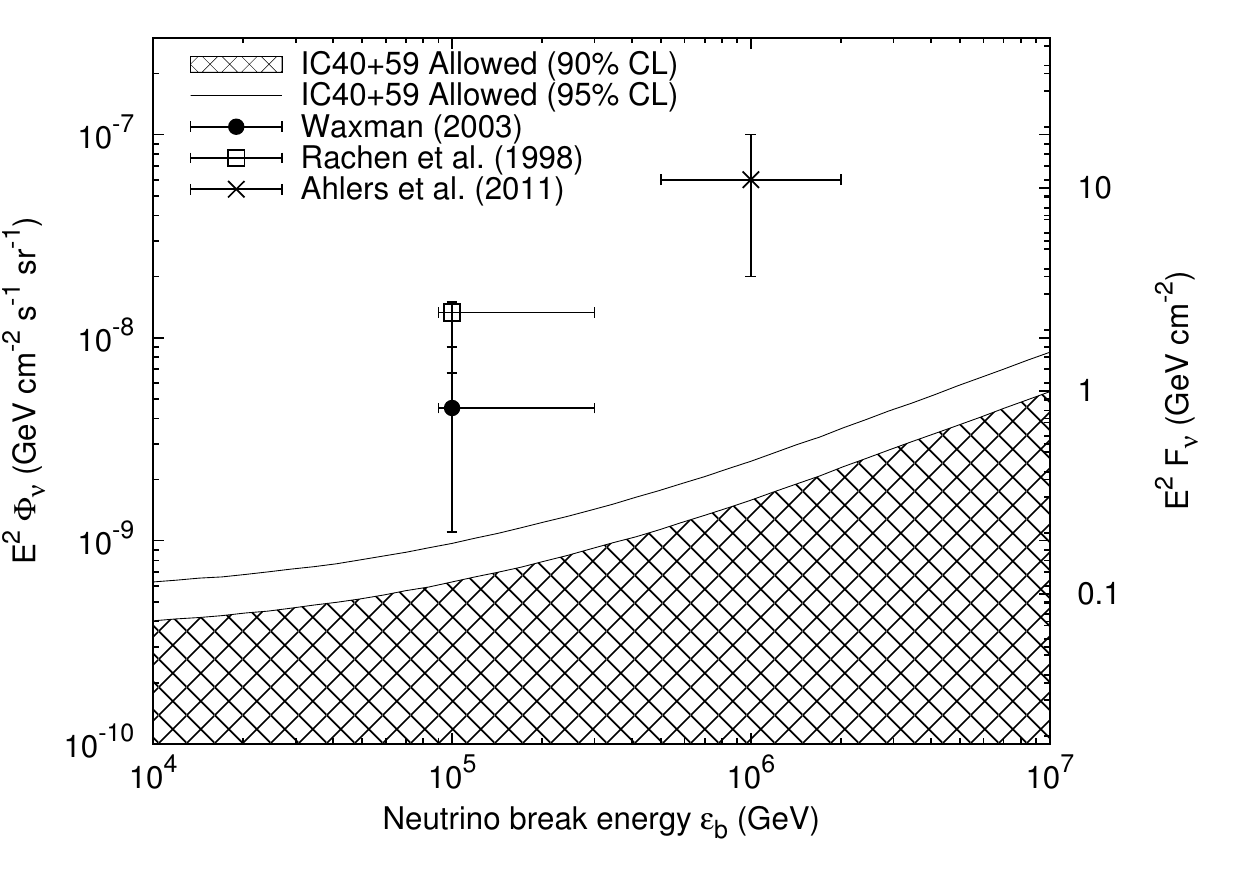}}
	\caption{Compatibility of some neutrino flux predictions based on cosmic ray production in GRBs with observations from IceCube. The cross-hatched area shows the 90\% confidence allowed values of the neutrino flux versus the neutrino break energy in comparison to model predictions with estimated uncertainties (points); the solid line shows the upper bound of the 95\% confidence allowed region. Taken from~\cite{nature}.}
	\label{fig:grbmodels}
\end{center}
\end{figure}

IceCube has performed a search for neutrinos in spatial and time coincidence with gamma ray bursts using the IC40 and IC59 data. During this period there were 215 GRBs in the northern sky used for a model dependent search~\cite{nature}. We also performed a model independent search over multiple time scales and including the southern sky bursts for the IC59 data set~\cite{nature}. For the model dependent analysis we detected no coincident neutrinos with an expectation for the model of 8.4 signal events. This gives a 90\% CL upper limit of 0.27 times the predicted flux (Figure~\ref{fig:grblimit}). Figure~\ref{fig:grbmodels} shows some alternative models along with their uncertainties, and compares them to the parameter range allowed by the IceCube model independent analysis. It is too early to exclude GRBs as sources of extragalactic cosmic rays. However, continuing results from IceCube in the upcoming years will either 1) observe neutrinos and have unambiguous evidence of hadronic acceleration, or 2) essentially rule out GRBs as the source of high energy cosmic rays. 

\subsection{Observation of Neutrino Oscillations}

Although IceCube's primary mission is the search for astrophysical neutrinos and discovering the source of cosmic rays, there are a number of physics topics accessible to the instrument including the measurement of neutrino properties using the large sample of atmospheric neutrinos recorded with the detector. As mentioned above, the completed detector records on the order of 50,000 well reconstructed high energy neutrinos per year depending on the event selection and energy range. In order to exploit this capability and to enhance the low energy performance of the detector for other physics (e.g. dark matter searches) part of the last two years of construction were used to infill the deep central region of the detector (DeepCore) with a higher density of DOMs. To highlight this enhanced capability we have performed an analysis to look for atmospheric muon neutrino oscillations using the partially completed DeepCore during the IC79 run. 

\begin{figure}
\begin{center}
	\resizebox{\linewidth}{!}{\includegraphics{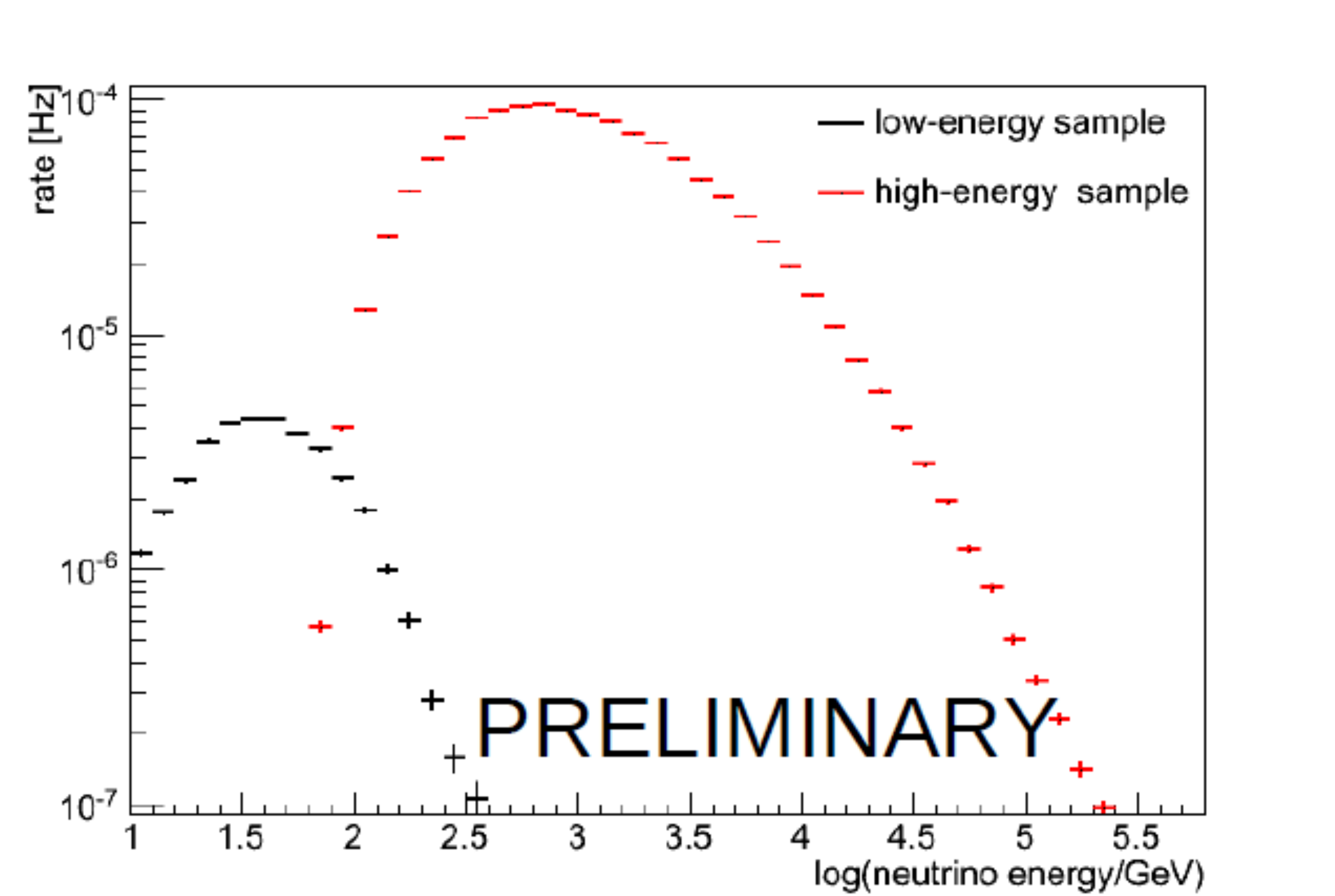}}
	\caption{Events versus neutrino energy for the high energy IceCube sample (red) and low energy DeepCore sample (black).}
	\label{fig:osc_energy}
\end{center}
\end{figure}

\begin{figure}
\begin{center}
	\resizebox{\linewidth}{!}{\includegraphics{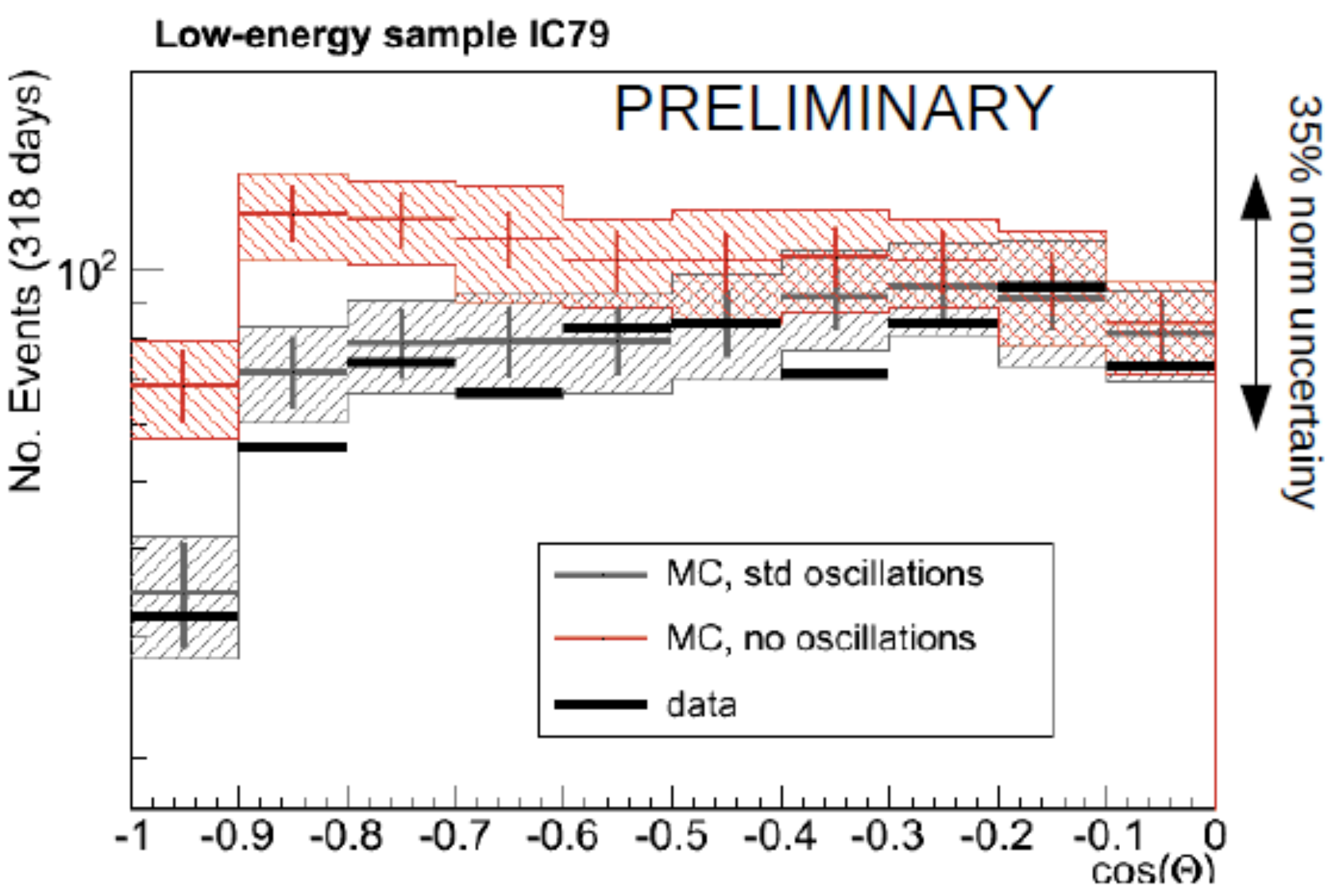}}
	\caption{Zenith angle distribution of low energy events.}
	\label{fig:osc_zen}
\end{center}
\end{figure}

\begin{figure}
\begin{center}
	\resizebox{\linewidth}{!}{\includegraphics{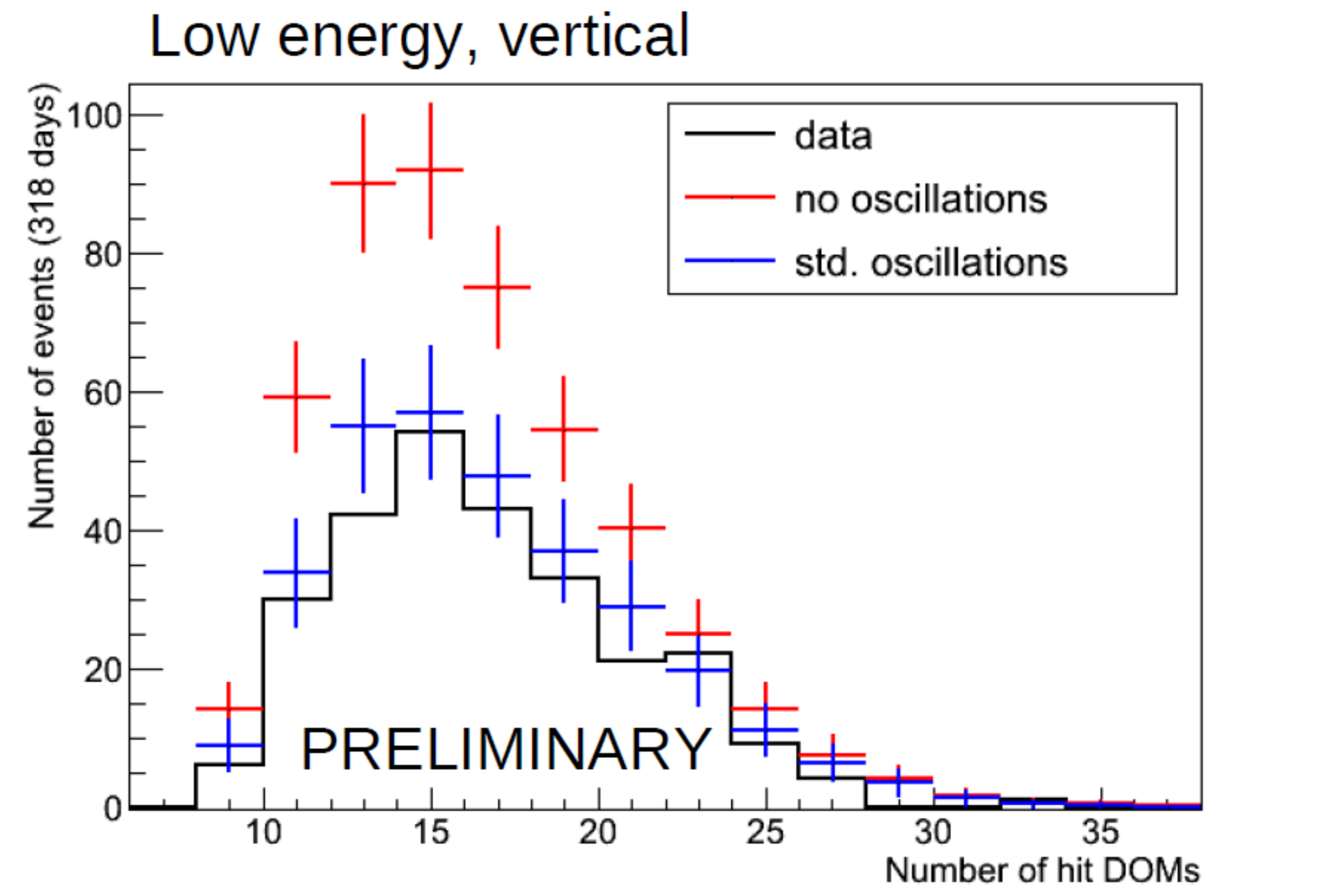}}
	\caption{n-channel distribution of directly upward going low-energy events.}
	\label{fig:osc_nch}
\end{center}
\end{figure}

This initial IC79 analysis is a simple extrapolation of basic cuts used at higher energies modified for the DeepCore. More sophisticated analyses and reconstructions are in development that will improve upon this initial demonstration. Shown in Figure~\ref{fig:osc_energy} are the energy distributions of muons from neutrinos contained in the DeepCore compared with the normal IceCube. These tight cuts yield 39,639 events in the high energy sample with a peak in the 300 GeV range, and 719 events in the low energy DeepCore sample with a peak in the 20-30 GeV range. The expected number of low energy events are 789 for standard oscillations and 1015 for no oscillations. Figure~\ref{fig:osc_zen} shows the zenith angle distribution for the low energy sample (solid bars) along with the predictions for standard oscillations (black crosses) and no oscillations (red crosses). The bands represent the bin-to-bin systematic error. There is a 35\% normalization error which is largely removed by simultaneously fitting to the high energy sample. The delta chi-squred between the data and the two distributions gives a p-value of $1 \times 10^{-8}$ for consistency with the no oscillation hypothesis. This fit used the zenith angle distribution only. Shown independently in Figure~\ref{fig:osc_nch} is the n-channel distribution (i.e number of DOM hits, which correlates with energy), and shows the data compared to the oscillation versus no oscillation hypothesis. 


This first  oscillation result from IceCube with DeepCore clearly demonstrates the ability to extend the sensitivity to lower energies with higher density instrumentation in the deep South Pole ice. We expect to be making measurements of the oscillation parameters in the near future. We have begun feasibility studies of extending the energy range to a few GeV with a very modest cost higher density upgrade with the goal of measuring the neutrino mass hierarchy~\cite{future}. Studies have indicated that if the appropriate energy and angular resolution can be achieved such a detector could measure the mass hierarchy with 4-12 sigma significance in five years~\cite{osc}.

\section{Conclusion and Future Outlook}
\label{sec:conclusion}

The IceCube detector completed construction in December 2010 and is now fully operational. The detector represents the first neutrino telescope to reach the cubic kilometer scale necessary for sensitivity of astrophysical interest. We have reported on results from the partially completed detector. Searches for diffuse, time-independent points sources and neutrino emission from GRBs have not yet found evidence for astrophysical emission. However, in the case of GRBs the result is already setting astrophysically important limits on the fireball and other models for GRBs and has caused a flurry of activity in more accurately modeling the expected neutrino emission. In the next years we expect to either see neutrinos from GRBs or essentially rule them out as the primary source of extragalactic cosmic rays. The point source searches will become sensitive to the generally expected fluxes in the coming 3-5 year of operation. 

We have demonstrated the capability of a large neutrino telescope detector to successfully extend the low energy response to the 10's of GeV region and observe neutrino oscillations. We are currently investigating whether a low cost and short time scale upgrade would be sufficient to further extend the response below 10 GeV with sufficient capability to measure the neutrino mass hierarchy in the next 5-7 years.

Although the search for diffuse astrophysical muon neutrinos has not found a significant excess, the best fit for the IC59 detector is non-zero and getting interesting. Shown in another talk from the IceCube experiment in these proceedings~\cite{HE} are two very high energy cascade events.  Taken by themselves or along with the slight excess in the muon channel, these events are suggestive that we may be on the threshold of discovering high energy astrophysical neutrinos bringing us into a new area in multi-messenger astronomy.

\section*{Acknowledgments}
I would like to thank Anne Schukraft, Andreas Gross, Nathan Whitehorn and Erik Blaufuss for help with material and for valuable discussion on the results.




\begin{thebibliography}{00}
\bibitem{history}
	A.~Roberts, Rev. Mod. Phys. {\bf 64}, 259 (1992)
\bibitem{halzen}
	F.~Halzen, Cosmic Neutrinos and the Energy Budget of Galactic and Extragalactic Cosmic Rays, 	arXiv:astro-ph/0604441v1 
\bibitem{WB}
	E.~Waxman and J.~Bahcall, Phys. Rev. D. {\bf 59}, 023002 (1998)
\bibitem{GRB}
	E.~Waxman and J.~Bahcall, High energy neutrinos from cosmological gamma-ray burst fireballs. Phys. Rev. Lett. {\bf 78}, 2292 (1997).
\bibitem{nature}
	R.~Abbasi et al., The IceCube Collaboration, Nature {\bf 84}, 351 (2012)
\bibitem{future}
	A.~Karle, these proceedings
\bibitem{osc}
	E.Kh.~Akhmedov, S.~Razzaque and A.Y.~Smirnov, arXiv:1205.7071 [hep-ph]
\bibitem{HE}
	A.~Ishihara, these proceedings
\end{thebibliography}




\end{document}